\documentclass[aps,pra,twocolumn,superscriptaddress,showpacs]{revtex4}
\usepackage{times}
\usepackage[dvips]{graphicx}

\begin{document}

\title{Conditional displacement operator for traveling fields}
\author{Simone Souza}
\affiliation{Instituto de F\'{\i}sica, Universidade Federal de Goi\'{a}s, 74.001-970, Goi%
\^{a}nia (GO), Brazil}
\author{A. T. Avelar}
\email[Electronic address:]{avelar@if.ufg.br}
\affiliation{Instituto de F\'{\i}sica, Universidade Federal de Goi\'{a}s, 74.001-970, Goi%
\^{a}nia (GO), Brazil}
\author{J. M. C. Malbouisson}
\affiliation{Instituto de F\'{\i}sica, Universidade Federal da Bahia, 40.210-310,
Salvador (BA), Brazil}
\author{B. Baseia}
\affiliation{Instituto de F\'{\i}sica, Universidade Federal de Goi\'{a}s, 74.001-970, Goi%
\^{a}nia (GO), Brazil}

\begin{abstract}
We show that the conditional displacement operator $\widehat{U}_{CD}=\exp [%
\hat{b}^{\dagger }\hat{b}(\beta \hat{a}^{\dagger }-\beta ^{\ast }\hat{a})]$
acting upon an arbitrary state of traveling waves can be well approximated
by the action of a Kerr medium placed between two beam splitters whose
respective second ports are fed by highly excited coherent states.
Applications to the generation of nonclassical states and
measurement of Wigner function of arbitrary states are also considered.
\end{abstract}

\maketitle

\section{Introduction}

The conditional displacement operator (CDO) has been extensively used
in the literature, e.g., by Milburn and Walls \cite{Milburn83} in quantum
nondemolition measurements via quantum counting; by Ban \cite{Ban94} in
theoretical studies of the photon statistics in the four-wave mixer; and by
Avelar \textit{et al.} \cite{Avelar05} in measurements of Wigner
characteristic function describing field states of running waves. In cavity
QED, Zou \textit{et al.} \cite{Zou04} have proposed the creation of the CDO
through a two-level atom interacting with a single-mode cavity field and
driven additionally by an external classical field. The authors used this
scheme for the measurements of the Wigner characteristic function. It has
been employed also for the generation of nonclassical states \cite%
{Weber07,Souza07}. However, to our knowledge, there is no suggestion on how
to implement this operator in a reliable way for arbitrary quantum states in
running waves.

Here we present a calculation which shows how to displace conditionally
arbitrary (pure or mixed) quantum states using a Kerr medium placed between
two beam splitters (along the path of the signal beam) whose respective
second ports are fed by appropriate, highly excited, coherent states. As
applications, we show how to engineer for running waves the even ($+$) and
odd ($-$) superpositions of an arbitrary single-mode state with its
displaced counterparts, $|\psi\rangle \pm \widehat{D}(\beta)|\psi \rangle$.
To this end the CDO device is coupled to one arm of a Mach-Zehnder
interferometer (MZI) fed by the vacuum and one-photon states, as applied for
Kerr medium by Sanders and Milburn \cite{Sanders89} to investigate
complementarity in a quantum nondemolition measurement. The present scheme
constitutes a modification of another one proposed by Villas-B\^{o}as 
\textit{et al.} \cite{Villas01} to measure directly the Wigner function. As
a by-product, it is shown that our scheme allows one to measure the Wigner
caracteristic function of the state $|\psi\rangle$ through the measurement
of photon-detection probabilities in the output of the MZI.

\section{Engineering the conditional displacement operator}

\begin{figure}[h]
\includegraphics[{height=2.5cm,width=6.5cm}]{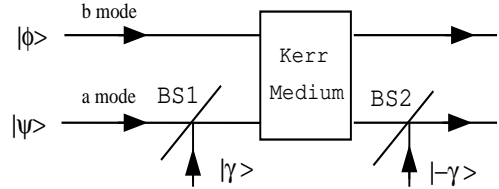}
\caption{Schematic illustration of the CDO device consisting of a
Kerr-medium between two beam-splitters along the path of the signal beam.}
\label{CDO}
\end{figure}

A schematic diagram of the procedure is shown in Fig.~\ref{CDO}. Both beam splitters
(BS1 and BS2) produce the action of the displacement operator $\widehat{D}%
_{a}(\alpha )=\exp (\alpha \hat{a}^{\dagger }-\alpha ^{\ast }\hat{a})$ on a
quantum state of the field-mode $a$, when the second port is fed by highly
excited coherent states $|\gamma \rangle $ and $|-\gamma \rangle $ as shown
in \cite{Paris96}. Thus, after the BS1 the state describing the whole
system becomes 
\begin{equation}
|\Psi ^{\prime }\rangle _{ab}=\widehat{D}_{a}(\alpha )|\psi \rangle
_{a}|\phi \rangle _{b}  \label{PsiLinha}
\end{equation}%
where $\alpha =R\gamma $, with $R\ll 1$ standing for the reflectance of the
BS1. 

The dispersive Kerr interaction between modes $a$ and $b$ is described by
the Hamiltonian \cite{Imoto85} 
\begin{equation}
\widehat{H}_{K}=\hbar K\hat{a}^{\dagger }\hat{a}\hat{b}^{\dagger }\hat{b}
\label{Hkerr}
\end{equation}
where $K$ is proportional to the third-order nonlinear susceptibility $\chi
^{(3)}$. So, the action of the Kerr medium upon (bipartite) field states is
represented by the unitary operator, 
\begin{equation}
\widehat{U}_{K}=exp(-i\theta \hat{a}^{\dagger }\hat{a}\hat{b}^{\dagger }\hat{%
b}),  \label{unitary}
\end{equation}
where $\theta =Kl/v$, $l$ is the length of the Kerr-medium and $v$ the
velocity of light in the medium. Due to the action of the Hamiltonian (\ref%
{Hkerr}) upon the modes $a$ and $b$, and the action of the BS2 corresponding
to a second displacement, the state of the system evolves to 
\begin{equation}
|\Psi ^{\prime \prime }\rangle _{ab}=\widehat{D}_{a}^{\dagger }(\alpha
)e^{-i\theta \hat{a}^{\dagger }\hat{a}\hat{b}^{\dagger }\hat{b}}\widehat{D}%
_{a}(\alpha )|\psi \rangle _{a}|\phi \rangle _{b}.
\end{equation}

Next, a little algebra furnishes 
\begin{eqnarray}
\widehat{D}_{a}^{\dagger }(\alpha )e^{-i\theta \hat{a}^{\dagger }\hat{a}\hat{%
b}^{\dagger }\hat{b}}\widehat{D}_{a}(\alpha ) &=&e^{-i\theta (\hat{a}%
^{\dagger }+\alpha ^{\ast })(\hat{a}+\alpha )\hat{b}^{\dagger }\hat{b}} 
\nonumber \\
&=&e^{(-i\theta \hat{a}^{\dagger }\hat{a}-i\theta \alpha \hat{a}^{\dagger
}-i\theta \alpha ^{\ast }\hat{a}-i\theta |\alpha |^{2})\hat{b}^{\dagger }%
\hat{b}}  \nonumber \\
&&
\end{eqnarray}%
and for realistic Kerr-media small values of phase shifts $\theta $ are produced in
laboratories; so, when adjusting the device to high values of $\alpha $ the foregoing
equation becomes 
\begin{equation}
\widehat{D}_{a}^{\dagger }(\alpha )e^{-i\theta \hat{a}^{\dagger }\hat{a}\hat{%
b}^{\dagger }\hat{b}}\widehat{D}_{a}(\alpha )\simeq e^{-i\theta |\alpha |^{2}%
\hat{b}^{\dagger }\hat{b}}\widehat{U}_{CD}(\beta ),  \label{aprox}
\end{equation}%
where $\beta =-i\theta \alpha $, with $|\beta |=\theta |\alpha |$ finite,
and 
\begin{equation}
\widehat{U}_{CD}(\beta )=\exp [\hat{b}^{\dagger }\hat{b}(\beta \hat{a}%
^{\dagger }-\beta ^{\ast }\hat{a})]  \label{UCDO}
\end{equation}%
is the wanted CDO. We emphasize that Eq.~(\ref{aprox}) is an algebraic operator relation,
and so, it does not depend on the input state of the CDO device.

\section{Applications}

As interesting applications of the foregoing scheme we will use it to
prepare the superposition states $|\psi \rangle \pm D( \beta )|\psi \rangle$
and to measure Wigner characteristic function of $|\psi \rangle$, where $%
|\psi \rangle$ is an arbitrary state incoming in the CDO.

\subsection{Engineering the superposed state: $|\protect\psi \rangle \pm D( 
\protect\beta )|\protect\psi \rangle $}

The superposed state of the kind $|\psi \rangle \pm D(\beta )|\psi \rangle $
in traveling waves has interesting applications in the literature. For
example, setting $|\psi \rangle =|\alpha \rangle $ and $\beta =-2\alpha $
one obtains the even ($+$) and odd ($-$) Schr\"{o}dinger's cat states $%
|\alpha \rangle \pm |-\alpha \rangle $; setting $|\psi \rangle =|0\rangle $
one obtains the superposition of the vacuum state and a coherent state, $%
|0\rangle \pm |\beta \rangle $, with the CDO playing the role of the optical
quantum switch \cite{Davidovich03} . Other interesting family of states can
be got from $|\psi \rangle =|n\rangle $, yielding the superposition $%
|n\rangle \pm D(\beta )|n\rangle $, which depends on the availability of a
Fock state $|n\rangle .$

\begin{figure}[h]
\includegraphics[{height=2.7cm,width=8.0cm}]{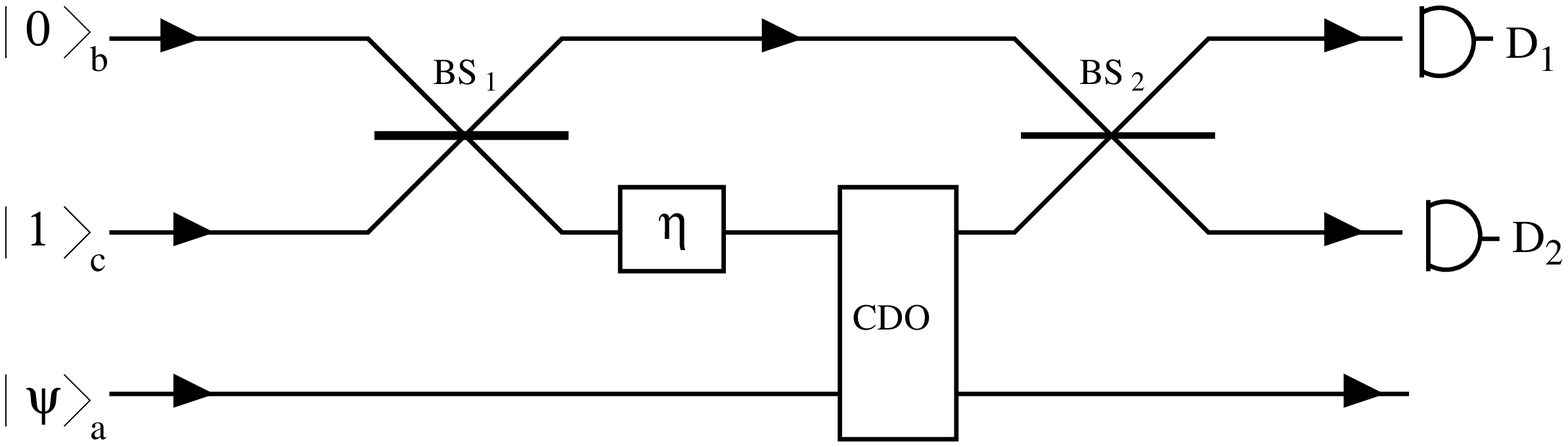}
\caption{Schematic illustration of the MZI, including a CDO device in one
arm, coupling the internal mode $b$ with the signal beam $a$.}
\label{Mach-Zehnder}
\end{figure}

To prepare the superposition $|\psi \rangle \pm D(\beta )|\psi \rangle $ one
needs a MZI associated with an auxiliary CDO and a phase shifter (PS), both
inserted in one arm (\textit{mode b}) of the MZI. The schematic setup is
depicted in the Fig.~\ref{Mach-Zehnder}. The CDO couples one of the internal
modes of the MZI (mode $b$) with an external mode (\textit{mode a}) where a
field in the state $|\psi \rangle $ is injected. Measurements of the
probability of photon detection in the output of the MZI allow us the
preparation of the superposition $|\psi \rangle \pm D(\beta )|\psi \rangle $
in the mode $a$, as follows. 

Consider the states $|0\rangle _{b}$ and $|1\rangle _{c}$ entering the ideal 
$50/50$ symmetric BS1 of the MZI, whose action upon them, such as that of
the BS2, is described by the following transformations \cite{Yeoman93} 
\begin{eqnarray}
|0\rangle _{b}|1\rangle _{c} &\rightarrow &(|0\rangle _{b}|1\rangle
_{c}+i|1\rangle _{b}|0\rangle _{c})/\sqrt{2}, \\
|1\rangle _{b}|0\rangle _{c} &\rightarrow &(|1\rangle _{b}|0\rangle
_{c}+i|0\rangle _{b}|1\rangle _{c})/\sqrt{2}.  \label{transformations}
\end{eqnarray}
The PS is assumed to add a phase $e^{i\eta }$ to the field crossing it.
Thus, just after the BS1 and the PS the (entangled) state of the whole
system reads, 
\begin{equation}
|\Psi \rangle _{abc}=\frac{1}{\sqrt{2}}\left( e^{i\eta }|0\rangle
_{b}|1\rangle _{c}+i|1\rangle _{b}|0\rangle _{c}\right) |\psi \rangle _{a} 
\text{ ,}
\end{equation}
where the state $|\psi \rangle _{a}$ in the second member stands for the
initial state entering the mode $a$.

Now, the action of the CDO device on the state $|\Psi \rangle _{abc}$ is
obtained from the relations 
\begin{eqnarray}
e^{-i\theta |\alpha |^{2}\hat{b}^{\dagger }\hat{b}}\widehat{U}_{CD}(\beta
)|\psi \rangle _{a}|0\rangle _{b} &=&|\psi \rangle _{a}|0\rangle _{b},
\label{U0} \\
e^{-i\theta |\alpha |^{2}\hat{b}^{\dagger }\hat{b}}\widehat{U}_{CD}(\beta
)|\psi \rangle _{a}|1\rangle _{b} &=&e^{-i\theta |\alpha |^{2}}{\widehat{D}}%
_{a}(\beta )|\psi \rangle _{a}|1\rangle _{b},  \nonumber \\
&&  \label{U1}
\end{eqnarray}%
and, after the BS2, the state of the whole system can be written as 
\begin{eqnarray}
|\Psi ^{\prime }\rangle _{abc} &=&\frac{1}{2}e^{-i\theta |\alpha
|^{2}}\left\{ |0\rangle _{b}|1\rangle _{c}\left[ e^{i\xi }-{\widehat{D}}%
_{a}(\beta )\right] \,|\psi \rangle _{a}\right.  \nonumber \\
&+&\left. i|1\rangle _{b}|0\rangle _{c}\left[ e^{i\xi }+{\widehat{D}}%
_{a}(\beta )\right] \,|\psi \rangle _{a}\right\} ,  \label{Psi3}
\end{eqnarray}%
with $\xi =\eta +\theta |\alpha |^{2}$. Note that in this scheme the PS
turns irrelevant the action of the additional factor $e^{-i\theta |\alpha
|^{2}\hat{b}^{\dagger }\hat{b}}$ accompanying the CDO in the Eq.~(\ref{aprox}%
).

At this point we can see that if the detector D1 fires (does not fire) while
D2 does not fire (fires) this corresponds to the output state $|1\rangle
_{b}|0\rangle _{c}$ ($|0\rangle _{b}|1\rangle _{c}$) in the BS2, and the mode%
$-a$ is projected onto the state 
\begin{equation}
|\psi ^{\pm }\rangle _{a}^{out}=\frac{1}{2}e^{-i\theta |\alpha |^{2}}\left[
e^{i\xi _{0}}\pm {\widehat{D}}_{a}(\beta )\right] \,|\psi \rangle _{a}.
\label{basiliana}
\end{equation}%
Finally, up to a global phase, our wanted state $|\psi \rangle \pm D(\beta
)|\psi \rangle $ is obtained from the Eq.~(\ref{basiliana}); this can be
achieved by adjusting the PS such that $\eta =-\theta |\alpha |^{2}$.

\subsection{Measuring the Wigner function}

As discussed previously in \cite{Avelar05}, a MZI supplemented by a
nonlinear medium allows us to measure the Wigner characteristic functions of
field states in traveling waves, yielding the reconstruction of the Wigner
functions themselves. The scheme in \cite{Avelar05} employs a MZI associated
with an auxiliary (specific) four-wave mixer that entangles the field state
in one arm of the apparatus with an external field crossing it. We now show
in which way the CDO implemented here can be used to measure the Wigner
function. To this end, we start from the success probabilities $P_{01}(\beta
,\xi _{0})$ and $P_{10}(\beta ,\xi _{0})$ to get the states given in the
Eq.~(\ref{basiliana}). Actually, in the present scenario these properties
furnish the values of Wigner characteristic function of the state $|\psi
\rangle _{a}$. As the photodetection experiment is repeated many times,
starting from the same initial field state $\hat{\rho}_{a}=|\psi \rangle
_{a}\langle \psi |_{a}$ , one obtains the probabilities 
\begin{eqnarray}
P_{01}(\beta ,\xi _{0}) &=&\frac{1}{2}-\frac{1}{2}\mathrm{Re}\left\{
e^{-i\xi _{0}}\mathrm{Tr}\left[ {\widehat{D}_{a}}(\beta )\hat{\rho}_{a}%
\right] \right\} ,  \label{proba} \\
P_{10}(\beta ,\xi _{0}) &=&\frac{1}{2}+\frac{1}{2}\mathrm{Re}\left\{
e^{-i\xi _{0}}\mathrm{Tr}\left[ {\widehat{D}_{a}}(\beta )\hat{\rho}_{a}%
\right] \right\} ,  \label{proba2}
\end{eqnarray}%
leading to 
\begin{eqnarray*}
\Delta P(\beta ,\xi _{0}) &=&P_{10}(\beta ,\xi _{0})-P_{01}(\beta ,\xi _{0})
\\
&=&\mathrm{Re[}e^{-i\xi _{0}}\chi _{a}(\beta )],
\end{eqnarray*}%
where the relation $\chi _{a}(\beta )=\mathrm{Tr}\left[ \hat{\rho _{a}}{%
\widehat{D}_{a}}(\beta )\right] $ has been employed \cite{Cahill69}. As
shown in \cite{Avelar05} the measurement of $\Delta P(\beta ,0)$ ($\Delta
P(\beta ,\pi /2)$) furnishes the real (imaginary) part of the characteristic
function. So, one gets 
\begin{equation}
\chi _{a}(\beta )=\Delta P(\beta ,0)+i\Delta P\left( \beta ,\pi /2\right) .
\end{equation}%
These two measurements actually lead to values of $\chi _{a}(\beta )$ at two
points (namely, $\beta $ and $-\beta $) owing to the property $\chi
_{a}^{\ast }(\beta )=\chi _{a}(-\beta )$.

Now, since the Wigner function is the Fourier transform of the
characteristic function $\chi _{a}$, namely, 
\begin{equation}
W(z)=\frac{1}{\pi ^{2}}\int d^{2}\beta \,\chi _{a}(\beta )\exp \left( z\beta
^{\ast }-z^{\ast }\beta \right) ,  \label{Wigner}
\end{equation}%
the determination of $\chi _{a}(\beta )$ for a reasonable set of values
permits the reconstruction of the Wigner function of the original state of
the external mode $a$. We note that $\beta $ is a free-parameter in the Eq.~(%
\ref{Wigner}) and, since $\beta =-i\theta \alpha $ where $|\alpha |$ was
assumed to be large and $\theta $ takes small values in the experiment, then 
$\beta $ can be varied in the range of interest through the control of the
parameters $\alpha $ and $\theta $.

\section{Conclusions}

In summary, this report introduces a method to engineer the conditional
displacement operator in running fields via a Kerr-medium placed between two
beam-splitters. This device, when conveniently inserted in one arm of a
Mach-Zehnder interferometer, allows one to prepare states of the kind $|\psi
\rangle \pm D(\beta )|\psi \rangle ,$ with the interesting applications
mentioned in the Sect. 2.A, and the measurement of their Wigner functions
(Sect.2.B). The procedure is also valid for mixed states. Concerning with
the reliability of the scheme, it is supported by recent technological
advances which have achieved photodetectors with efficiency near $100\%$ 
\cite{Imamoglu02} and stable single-photon sources yielding many single
photons, e.g., using: single quantum dots \cite{Michler00}; single molecules 
\cite{Lounis00}; diamond colour centers \cite{Beveratus02}; atoms \cite%
{Kuhn00}; turnstile devices \cite{Kim99}; and also traditional parametric
down conversion methods \cite{PDC}. With respect to the Wigner function, it
is worth mentioning a pertinent comparison between the present scheme 
and the Ref.~\cite{Villas01}: in \cite{Villas01} the Wigner function
is measured directly and in this aspect it is advantageous. However, \cite%
{Villas01} requires a very large nonlinear susceptibility yielding $%
\theta =\pi $, whose experimental implementation is very hard till now. 
Here, our goal is achieved via the use of small values of
phase-shifts $\theta $, available in laboratories through realistic
Kerr-medium, which constitutes a remarkable result from the experimental
point of view, a feature having no guarantee in Ref. \cite{Avelar05}.
In fact, following various trials to get large Kerr nonlinearities \cite%
{Kerr}, a recent result by Munro \textit{et al.} \cite{Munro05} has achieved
phase shifts $\theta $ $\simeq 0.01$, with residual absorption rates less
than $1\%$. Finally, we mention that decoherence effects, not
considered here, could be included via the phenomenological operator
approach, as implemented in \cite{Serra02}.

We thank the CAPES and CNPq, Brazilian agencies, for partial supports.

\end{document}